# Effect of SiC-Impurity Layer and Growth Temperature on MgB$_2$ Superconducting Tapes Fabricated by HPCVD[**]

By *Mahipal Ranot, Won Kyung Seong, Soon-Gil Jung, Won Nam Kang[*], J Joo, C-J Kim, B-H Jun, and S Oh*

The influence of SiC-impurity layer and growth temperature on microstructure and superconducting properties were studied for MgB$_2$ superconducting tapes. The pulsed laser deposition (PLD) system was used for the deposition of amorphous SiC-impurity layers on the flexible metallic Cu (001) tapes. The MgB$_2$ superconducting tapes were fabricated by growing MgB$_2$ films on the top of SiC/Cu tapes over a wide temperature range of 460 – 600 °C by using hybrid physical-chemical vapor deposition (HPCVD) system. Among all tapes, the MgB$_2$/SiC/Cu tape deposited at a temperature of 540 °C has the highest $T_c$ of ~ 37.7 K. Scanning electron microscopy (SEM) images revealed the hexagonal shaped MgB$_2$ grains with good connectivity, and their sizes were found to vary with growth temperatures. As compared to MgB$_2$/Cu tapes, the MgB$_2$/SiC/Cu tapes exhibited opposite trend in the dependence of critical current density ($J_c$) with deposition temperatures. The improved $J_c(H)$ behavior could be explained on the basis of the enhanced flux pinning force density ($F_p$) for MgB$_2$/SiC/Cu tapes upon increasing growth temperature.

[*]   Prof. W. N. Kang, M. Ranot, Dr. W. K. Seong, S.-G. Jung
      BK21 Physics Division and Department of Physics, Sungkyunkwan University, Suwon 440-746 (Korea)
      Email: wnkang@skku.edu
      Prof. J. Joo
      School of Advanced Materials Science and Engineering, Sungkyunkwan University, Suwon 440-746 (Korea)
      Dr. C.-J. Kim, Dr. B.-H. Jun
      Neutron Science Division, Korea Atomic Energy Research Institute, Daejeon 305-353, (Korea)
      Dr. S. Oh
      Material Research Team, National Fusion Research Institute, Daejeon 305-333, (Korea)

[**]  This work was supported by Mid-career Researcher Program through National Research Foundation of Korea (NRF) grant funded by the Ministry of Education, Science & Technology (MEST) (No. 2010-0029136) and the Global Partnership Program through the NRF funded by the MEST (M60602000012).

**Short summary**

We report on the effect of SiC-impurity layer as well as growth temperature on the superconducting properties of MgB$_2$ tapes. Firstly, SiC-impurity layers were grown on Cu (001) tapes by pulsed laser deposition system. Subsequently, MgB$_2$ films were fabricated on SiC/Cu tapes over a wide temperature range of 460 – 600 °C by using hybrid physical-chemical vapor deposition system. All the prepared tapes were characterized by XRD, SEM, resistivity and magnetization measurements. The MgB$_2$/SiC/Cu tapes showed improved flux pinning and enhanced critical current density ($J_c$) than the MgB$_2$/Cu tapes. Our results demonstrate that HPCVD is a promising technique to fabricate thick MgB$_2$ superconducting tapes with high $J_c$ values for large scale applications.

## 1. Introduction

For large scale electric power applications, the superconducting $MgB_2$ ($T_c \sim 39$ K) tapes and wires are required to be produced in long lengths to replace the conventional metallic superconductors, such as Nb-Ti ($T_c = 9$ K) and $Nb_3Sn$ ($T_c = 18$ K).[1] The commonly used method for the fabrication of $MgB_2$ tape and wire is the so-called powder-in-tube (PIT) process,[2,3] and the encouraging results were reported for PIT-processed conductors.[4] However, the main problem of the PIT-processed tapes and wires is the low core density of $MgB_2$ (below ~ 50%). On the other hand, the highly dense $MgB_2$ phase can be easily obtained by growing $MgB_2$ on the metallic substrates with the so-called coated-conductor (CC) approach.[5] The CC approach has been widely used for high-$T_c$ cuprate superconductors, $YBa_2Cu_3O_{7-\delta}$ (Y-123) compound,[6] for example, where Y-123 films are deposited on various buffer layers on a metallic substrate. The highly textured Y-123 is required for high critical current density ($J_c$) in Y-123 CCs due to the poor superconducting coupling across the grain boundaries of this material.[7] In contrast to the cuprates high-$T_c$ superconductors, the grain boundaries do not limit the current flow of $MgB_2$ and indeed act as effective flux-pinning centers, which enhance the $J_c$.[8] Therefore, the direct deposition of $MgB_2$ onto the metallic substrate is possible and can avoid the necessity of buffer layers to reduce the cost as well as the complexity of the overall deposition process.

The hybrid physical–chemical vapor deposition (HPCVD) has been proved as the most effective technique for the fabrication of high-quality $MgB_2$ films.[9,10] In addition, strong enhancement of critical current density and record high values of upper critical field ($H_{c2}$) over 60 T have been reported for carbon-doped $MgB_2$ films produced by HPCVD.[11,12] Therefore, the impact of this technique for the fabrication of $MgB_2$ tapes and wires is of great interest. In our previous study, we have deposited $MgB_2$ on Cu (001) tapes by HPCVD, where the $J_c$ is decreased with an increasing external magnetic field because of their poor flux pinning properties.[13] Therefore, it is imperative to improve further the flux pinning properties of $MgB_2$ in the magnetic fields to make it useful for practical applications. In this work, we used SiC layer which is grown on metallic Cu (001) tape as an impurity for creating additional pinning sites in $MgB_2$. The amorphous layer of SiC can be act as an impurity for creating additional defects in $MgB_2$, because the binding energy of an amorphous phase is lower than that of a crystalline phase. Moreover, the critical properties, such as $T_c$, $J_c$, and $H_{c2}$ of superconductors are strongly influenced by heat treatment conditions. Therefore, the performance of any superconductor can be improved further by optimizing the growth temperature. Thus, it is necessary to investigate the heat treatment effect on the $MgB_2$/SiC/Cu tapes. This paper discusses the changes in microstructure and superconducting properties of $MgB_2$/SiC/Cu tapes fabricated within a temperature range of 460 – 600 ºC.

## 2. Results and Discussion

The X-ray $\theta$–$2\theta$ scan of textured Cu (001) tape, $MgB_2$/Cu tape fabricated at 460 °C, and $MgB_2$/SiC/Cu tapes prepared at different temperatures of 460, 500, 540, and 600 °C are shown in Figure 1. The XRD patterns showed only (000$l$) peaks of $MgB_2$ for $MgB_2$/Cu and SiC-coated tapes except for the secondary phases. In $MgB_2$/Cu tape, $MgCu_2$ and $Mg_2Cu$ are the main secondary phases, whereas in $MgB_2$/SiC/Cu tapes only $MgCu_2$ is observed as a major secondary phase. These secondary phases are observed because Mg is highly reactive with Cu tape during deposition process. Moreover, there is no indication of the presence of SiC and $Mg_2Si$ compounds which were observed in previously reported SiC-doped $MgB_2$ tapes and wires.[14,15] We anticipate that SiC particles of SiC-impurity layer are diffused into the $MgB_2$ matrix as defects during the film growth process, as no SiC layer was remained there between Cu tape and $MgB_2$ layer. As the sizes of the SiC diffused particles are small this is hard to detect in XRD. We were also unable to find any shift in the $MgB_2$ peaks for $MgB_2$/SiC/Cu tapes, since C substitutions into B sites leads to a peak shift,[16] which means that there are hardly any chemical reactions between Mg and SiC-impurity layers. The reason why SiC did not decompose and not react with Mg is most probably the low growth temperatures.

The normalized resistivity ($\rho/\rho_{40K}$) versus temperature curves for $MgB_2$/Cu and $MgB_2$/SiC/Cu tapes are plotted in Figure 2. All tapes exhibited the critical temperatures ranging between 36 and 38 K with superconducting transition width ($\Delta T_c$) of about 0.4 – 1 K. The SiC-coated MgB$_2$ tapes have higher $T_c$ than that of MgB$_2$/Cu tape. Among all the tapes, the tape fabricated at a temperature of 540 °C has the highest $T_c$ of ~37.7 K with a broad $\Delta T_c$ ~1.1 K which indicating non-homogeneity of the sample. We believe that the increase in $T_c$ is most probably due to improvement in crystallinity with the increase of growth temperature, as we observed in our previous report,[13] and it is not due to the SiC-impurity addition. Because the doping of carbon or carbon compounds in MgB$_2$ generally decreases the $T_c$ value due to C substitution into the B site [16,17] these results indicate that the SiC-impurity layer does not decompose into C, Si and Mg$_2$Si, and there is no substitution of C into the B site of MgB$_2$.

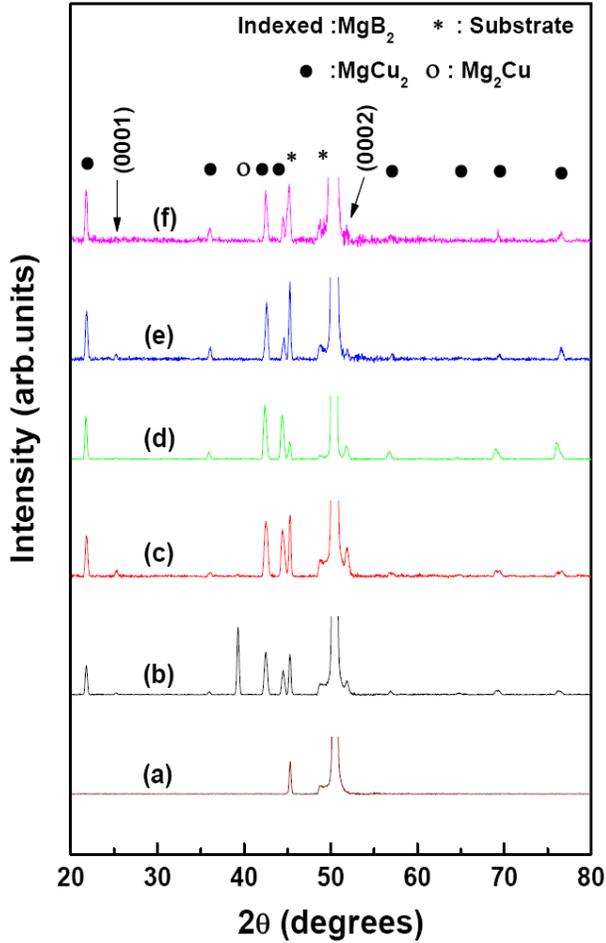

Fig. 1. XRD patterns for Cu (001) tape (a), MgB$_2$/Cu tape grown at 460 °C (b), and MgB$_2$/SiC/Cu tapes fabricated at various temperatures of 460 °C (c), 500 °C (d), 540 °C (e) and 600 °C (f).

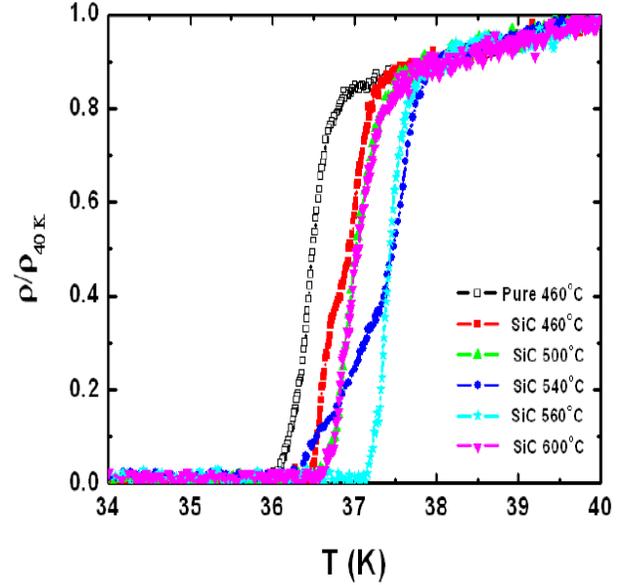

Fig. 2. The normalized resistivity ($\rho/\rho_{40\ K}$) as a function of temperature for MgB$_2$/Cu and MgB$_2$/SiC/Cu tapes deposited at various temperatures of 460, 500, 540, 560, and 600 °C.

The thickness of the SiC-impurity layer of 20 nm was measured by focused ion beam (FIB) technique. The surface morphologies and the thicknesses of MgB$_2$/Cu and MgB$_2$/SiC/Cu tapes are measured by scanning electron microscopy (SEM). The thickness of all the fabricated MgB$_2$ tapes was ~2 $\mu$m. Figure 3 shows the surface morphologies for MgB$_2$/Cu (a) and MgB$_2$/SiC/Cu tapes fabricated at 460 °C (b), 500 °C (c), 540 °C (d), 560 °C (e) and 600 °C (f). It is seen from Figure 3 that the MgB$_2$ grains of all tapes are well connected. For the MgB$_2$/Cu tape grown at 460 °C, the average grain size is estimated to be 350 nm, while for the MgB$_2$/SiC/Cu tape prepared at same temperature the average grain size is 525 nm. It seems there is less reactivity between Mg and Cu for SiC-coated tape, and therefore enough Mg is available to react with B to form MgB$_2$ phase. Moreover, in our previous work, the MgB$_2$ grain size was observed to be depending on the thickness of SiC-impurity layer.[18] With increasing the growth temperature average grain size was found to increase and reached its maximum value of 850 nm at a temperature of 500 °C, after that it starts to decrease. The suppression of MgB$_2$ grain growth above temperature of 500 °C could be due to the relatively low sticking coefficient of Mg at higher temperatures.[19] The samples fabricated at temperatures above 540 °C have dense microstructure but the estimation of grain size seems difficult. In the SEM cross-sectional images (not shown here) of MgB$_2$/SiC/Cu tapes, the SiC layer was not observed between Cu tape and MgB$_2$ layer. We anticipate that the amorphous SiC layer was diffused into the MgB$_2$ matrix as SiC particles during MgB$_2$ film growth process.

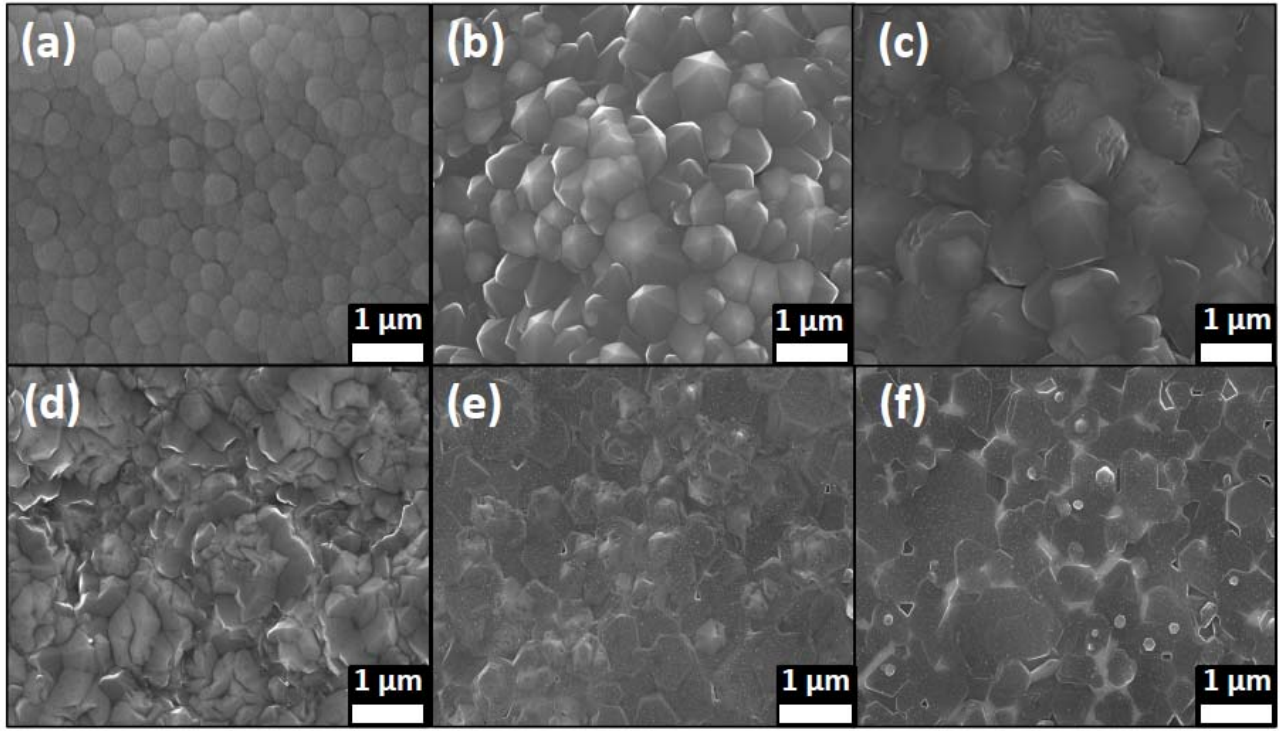

Fig. 3. SEM images of MgB$_2$/Cu tape grown at 460 °C (a) and of MgB$_2$/SiC/Cu tapes prepared at 460 °C (b), 500 °C (c), 540 °C (d), 560 °C (e) and 600 °C (f). The hexagonal shaped MgB$_2$ grains with good connectivity were observed among themselves.

The critical current density ($J_c$) is estimated from the magnetization hysteresis ($M - H$) loops using the Bean's critical state model. The $J_c(H)$ curves for MgB$_2$/Cu and MgB$_2$/SiC/Cu tapes as a function of growth temperature measured at 5 and 20 K are shown in Figure 4a) and b), respectively. We can see that both at 5 and 20 K the self-field $J_c$ for SiC-coated MgB$_2$ tapes is higher than that of MgB$_2$/Cu tape. It is well established that SiC doping in MgB$_2$ using the conventional *in situ* technique shows a decrease in $T_c$ and reduction of the self-field $J_c$ due to the high level of C substitution on the B site.[15] Recently Li *et al.* report the enhancement of self-field $J_c$ for SiC-MgB$_2$ composite made by the diffusion process without the observation of any chemical reaction between SiC and MgB$_2$.[20] No reduction of the $T_c$ and the absence of any chemical reaction between SiC-impurity layer and MgB$_2$ might be the reason for improvement in the self-field $J_c$ of our MgB$_2$/SiC/Cu tapes. Furthermore, on increasing the deposition temperature an increase in $J_c$ under applied field was observed for SiC-coated tapes. The optimal growth temperature for the strong improvement in $J_c$ was observed to be 540 °C. These results are in contrast to our previously reported data for MgB$_2$/Cu tapes, where MgB$_2$ films were directly deposited on the textured Cu tapes and they showed a decrease in $J_c$ with increasing growth temperature.[13] The enhanced $J_c$ for MgB$_2$/SiC/Cu tapes could be due to the improved flux pinning by grain boundaries along with the additional defects created by SiC-impurity layer which act as effective flux pinning centers.[17] It is also noted that the in-field $J_c$ of MgB$_2$/Cu tape around 4 and 2 T at 5 and 20 K, respectively is comparable with the MgB$_2$/SiC/Cu tapes. This might be due to the smaller grain size of 350 nm for MgB$_2$/Cu tape and as a result this sample has a high density of grain boundaries which are known to be the main pinning source in MgB$_2$ superconductor.[8,21]

Figure 5 shows the magnetic field dependences of the flux pinning force densities ($F_p$) at 5 K for MgB$_2$/Cu and MgB$_2$/SiC/Cu tapes grown at various temperatures, the $F_p$ is derived from the $J_c(H)$ data of Figure 4a). The $F_p - H$ graph clearly shows an increase in the flux pinning force density with increasing growth temperature for SiC-coated MgB$_2$ tapes as compared to MgB$_2$/Cu tape. The largest enhancement of $F_p$ is observed for the MgB$_2$/SiC/Cu tape fabricated at 540 °C which has a maximum $F_p$ value about 5.4 times higher than the MgB$_2$/Cu tape. It indicates that the MgB$_2$/SiC/Cu tapes have higher density of pinning centers, such as grain boundaries and additional defects than MgB$_2$/Cu tape as a consequence of SiC-impurity layer.

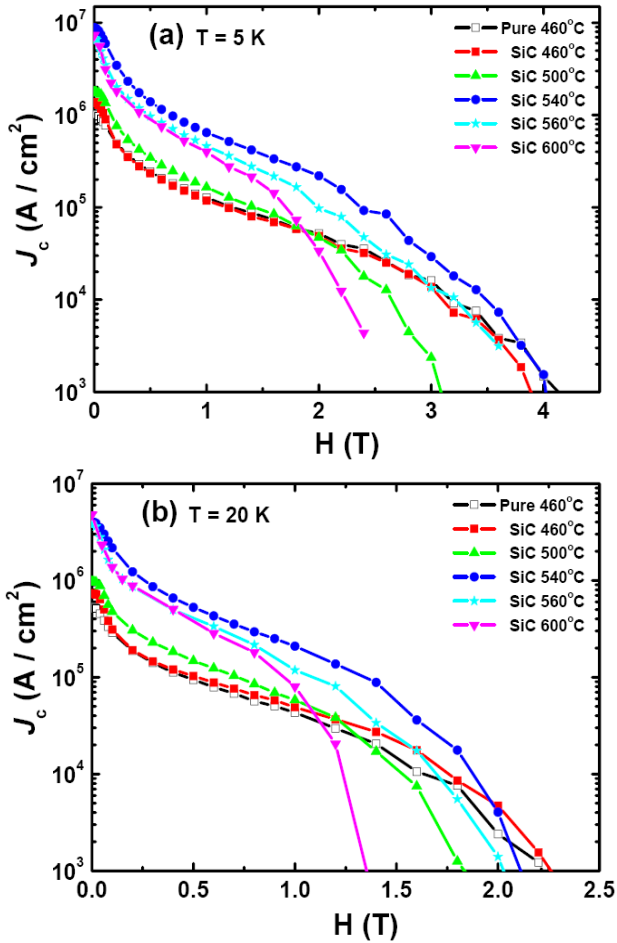

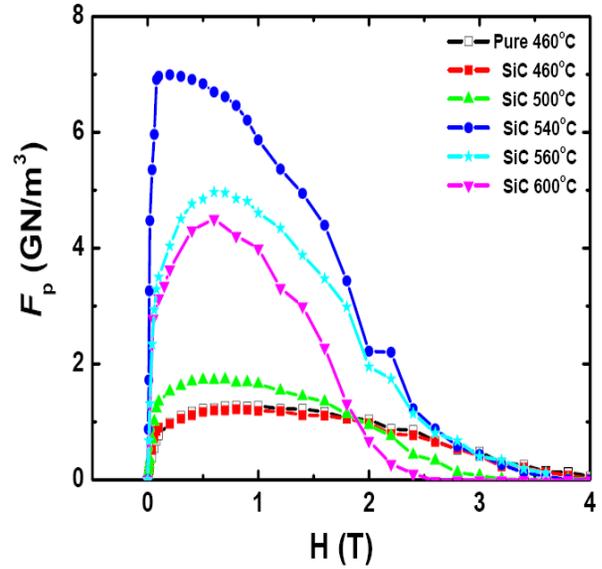

Fig. 4. The $J_c$ ($H$) curves as a function of growth temperature for MgB$_2$/Cu and MgB$_2$/SiC/Cu tapes measured at 5 K (a) and 20 K (b). The optimal growth temperature for the strong improvement in $J_c$ for MgB$_2$/SiC/Cu tapes was observed to be 540 °C.

## 3. Conclusions

We have investigated the effect of SiC-impurity layer and growth temperature on microstructure and superconducting properties for MgB$_2$ superconduc-ting tapes fabricated over a wide temperature range of 460 – 600 °C by using HPCVD. The XRD patterns showed only (000$l$) peaks of MgB$_2$ for MgB$_2$/Cu and MgB$_2$/SiC/Cu tapes. The SEM micrographs revealed that the average MgB$_2$ grain size changes with the deposition temperature. An increase in the flux pinning force density values upon increasing growth temperature was observed for the MgB$_2$/SiC/Cu tapes. In contrast to MgB$_2$/Cu tapes, the MgB$_2$ on SiC-coated Cu tapes exhibited opposite trend in the dependence of $J_c$ with growth temperature. The optimal growth temperature for the strong improvement in $J_c$ was observed to be 540 °C. The improved flux pinning by the additional defects created by SiC-impurity layer along with the MgB$_2$ grain boundaries lead to strong improvement in $J_c$ for the MgB$_2$/SiC/Cu tapes. These results imply that the HPCVD is a promising technique to fabricate MgB$_2$ superconducting tapes with high $J_c$ values for practical applications.

Fig. 5. Magnetic field dependence of the flux pinning force density ($F_p$) at 5 K for MgB$_2$/Cu and MgB$_2$/SiC/Cu tapes grown at various temperatures.

## 4. Experimental

*Cu Tape Preparation*: The textured Cu (001) tapes have been prepared by one hour heat treatment of Cu ingot in a tube furnace at 900 °C under a flow of a mixture of argon and hydrogen (5%) gases.[22] It is well known that, the columnar grains growth is crucial for obtaining high $J_c$ values in MgB$_2$. Therefore, we used textured Cu (001) tapes to induce *c*-axis oriented MgB$_2$ grains growth with columnar structure. The thickness of as-prepared Cu tapes was ~ 90 $\mu$m and the tapes were cut into sizes of 1 cm × 1 cm.

*SiC-Impurity Layer Deposition*: The pulsed laser deposition (PLD) system was used for the deposition of SiC-impurity layers. The amorphous SiC-impurity layers of thickness 20 nm were deposited on the textured Cu (001) tapes by PLD at room temperature with a background pressure of ~ $10^{-6}$ Torr. Laser beams were generated using a Lambda Physik KrF excimer laser ($\lambda$ = 248 nm). The pulse energy was set at 300 mJ with a repetition rate of 8 Hz. The SiC target (99.99%) was used for the deposition of impurity layers. The PLD system used in this study is described in more detail.[23] After depositing the SiC-impurity layers on the Cu tapes, we used HPCVD system for the fabrication of MgB$_2$ superconducting tapes by growing MgB$_2$ films on SiC/Cu tapes. The amorphous layer of SiC may prevent further orientation effect of Cu-substrate on MgB$_2$ layer. But we found that this layer did not affect the orientation effect of Cu-substrate on MgB$_2$ layer.

*HPCVD system*: The HPCVD system consists of a vertical quartz tube reactor having a bell shape (inner diameter, 65 mm; height, 200 mm), inductively coupled heater, and a load-rock chamber. A schematic diagram of this system has previously been described in detail.[10] The HPCVD has been the most effective technique for depositing high-quality MgB$_2$ films.[9] The small Mg chips (99.999%) (diameter, 2 - 3 mm) and B$_2$H$_6$ gas (5% in H$_2$ gas) were used as Mg and B sources, respectively.

*MgB$_2$ Film Growth Mechanism*: The growth mechanism of MgB$_2$ film followed the Volmer-Weber growth mode. In the Volmer-Weber growth mode, the smallest stable clusters nucleate on the substrate and grow in 3D to form islands. At the beginning of the growth, islands of hexagonal shaped MgB$_2$ form on the substrate, which grow and coalesce into a continuous film at larger film thickness. When the islands coalesce, they ''zip up'' because the surface energy of the islands is larger than the free energy of the grain boundaries. The more detail on the deposition technique and mechanism can be found in.[24]

*Fabrication of MgB$_2$ Superconducting Tapes*: The MgB$_2$ superconducting tapes were fabricated by using the HPCVD process. In this process, the Cu tape coated with SiC-impurity layer was placed on the top surface of a susceptor and Mg chips were placed around it. For the deposition of MgB$_2$ film on a SiC/Cu tape, the reactor was firstly evacuated to a base pressure of ~ $10^{-3}$ Torr using rotary pump and purged several times by flowing high purity argon and hydrogen gases. Prior to the film growth, the susceptor along with SiC/Cu tape and Mg chips were inductively heated towards the set temperature under a reactor pressure of 50 Torr in atmosphere of H$_2$. Upon reaching the set temperature, a boron precursor gas, B$_2$H$_6$ (5% in H$_2$) gas was introduced into the reactor to initiate the film growth. The flow rates were 90 sccm for the H$_2$ carrier gas and 10 sccm for the B$_2$H$_6$/H$_2$ mixture. Finally, the fabricated MgB$_2$/SiC/Cu tape was cooled down to room temperature in a flowing H$_2$ carrier gas. The MgB$_2$/SiC/Cu tapes were fabricated in the temperature range of 460–600 °C. A MgB$_2$/Cu tape was also prepared at a temperature of 460 °C for comparison. The thickness of all the fabricated MgB$_2$ tapes was ~ 2 $\mu$m with a growth rate of 0.2 $\mu$m min$^{-1}$.

*Tape Characterization*: The crystal structures of MgB$_2$/Cu and MgB$_2$/SiC/Cu tapes were investigated by XRD (D8 discover, Bruker AXS) using Cu K$\alpha$ as an X-ray source. The thickness of the SiC-impurity layer was measured by focused ion beam (FIB) technique. SEM (JEOL, JSM – 7000F) was used for measuring the surface morphologies and the thicknesses of the tapes. The standard four-probe method was used in order to measure the temperature dependence of resistivity for all the prepared tapes. The magnetization hysteresis ($M - H$) measurements were carried out on all the tapes using a magnetic property measurement system (XL-5S, Quantum Design).